\documentclass[prd,preprint,tightenlines,floatfix,showpacs,preprintnumbers,nofootinbib,eqsecnum]{revtex4}

 \usepackage[dvips,final]{graphicx}
  \usepackage{amssymb}
   \usepackage{amsmath}
    \usepackage{amsfonts}
     \usepackage{epsfig}
      \usepackage{bm}

\newcommand{\GeV}{\textrm{GeV}}

 \def\Pom{{ I\!\!P}}
 \def\Reg{{ I\!\!R}}

\begin{document}

\begin{flushright}
LU TP 11-32\\
September 2011
\end{flushright}

\title{Drell-Yan diffraction: breakdown of QCD factorisation}

\author{R. S. Pasechnik}
 \email{Roman.Pasechnik@thep.lu.se}
 \affiliation{
Theoretical High Energy Physics, Department of Astronomy and
Theoretical Physics, Lund University, S\"olvegatan 14A, SE 223-62
Lund, Sweden}

\author{B. Z. Kopeliovich}
 \email{Boris.Kopeliovich@usm.cl}
 \affiliation{ Departamento de F\'{\i}sica Universidad T\'ecnica
 Federico Santa Mar\'{\i}a; and\\
 Instituto de Estudios Avanzados en Ciencias e Ingenier\'{\i}a; and\\
 Centro Cient\'ifico-Tecnol\'ogico de Valpara\'iso;\\
 Casilla 110-V, Valpara\'iso, Chile}

\begin{abstract}
We consider the diffractive Drell-Yan process in proton-(anti)proton
collisions at high energies in the color dipole approach. The
calculations are performed at forward rapidities of the leptonic
pair. Effect of eikonalization of the universal ``bare''
dipole-target elastic amplitude in the saturation regime takes into
account the principal part of the gap survival probability. We
present predictions for the total and differential cross sections of
the single diffractive lepton pair production at RHIC and LHC
energies. We analyze implications of the QCD factorisation breakdown
in the diffractive Drell-Yan process, which is caused by a specific
interplay of the soft and hard interactions, and resulting in rather
unusual properties of the corresponding observables.
\end{abstract}


\maketitle

\section{Introduction}


The exclusive diffractive production of particles in hadron-hadron
scattering at high energies is one of the basic tools for both
experimental and theoretical studies of the small-$x$ and
nonperturbative QCD physics. The understanding of the mechanisms of
inelastic diffraction came with the pioneering works of Glauber
\cite{Glauber}, Feinberg and Pomeranchuk \cite{FP56}, Good and
Walker \cite{GW}. If the incoming plane wave contains components
interacting differently with the target, the outgoing wave will have
a different composition, i.e. besides elastic scattering a new {\it
diffractive} state will be created (for a detailed review on QCD
diffraction, see Ref.~\cite{KPSdiff}). Among the most important
examples, the leading twist diffractive Drell-Yan (DDY) process is
of a special interest since it gives rise to a clean experimental
signature for the QCD factorisation breaking effects where soft and
hard interactions interplay with each other \cite{KPST06}, thus,
providing an access to the soft QCD physics \cite{RP11}.

Typically, the single-diffractive Drell-Yan reaction in $pp$ collisions is
characterized by a relatively small momentum transfer between the colliding protons,
such that one of them, e.g.
$p_1$, radiates a hard virtual
photon $k^2=M^2\gg m_p^2$ and hadronizes into a hadronic
system $X$ both moving in forward direction and
separated by a large rapidity gap from the second
proton $p_2$, which remains intact, i.e.
\begin{eqnarray}
p_1+p_2\to \gamma^*(l^+l^-)+X+(gap)+p_2
\end{eqnarray}
Both the di-lepton and $X$, the debris of $p_1$, stay in the forward
fragmentation region. In this case, the virtual photon is
predominantly emitted by the valence quarks of the proton $p_1$.
Below we will refer to this as the diffractive Drell-Yan process
at forward rapidities.
Notice that this is different from double diffractive
Drell-Yan process, where the di-lepton $l^+l^-$ is produced at
central rapidities, while both protons survive the collision (see e.g.
Ref.~\cite{Antoni11}).
Then, the $\gamma^*$ can be emitted by a sea quark or antiquark.
We postpone this case for future
studies, and  concentrate here on the single diffractive Drell-Yan process.

In some of previous studies Refs.~\cite{Antoni11,Landshoff-DDY} of the single diffractive Drell-Yan reaction
the analysis  was made within the phenomenological Pomeron-Pomeron
and $\gamma$-Pomeron fusion mechanisms using the Ingelman-Shlein
approach \cite{ISh} based on QCD and Regge factorization. This led to specific features of the differential cross sections similar to
those in diffractive DIS process, e.g., a slow increase of the
diffractive-to-inclusive DY cross sections ratio with c.m.s. energy
$\sqrt{s}$, its practical independence on the hard scale, the
invariant mass of the lepton pair squared, $M^2$ \cite{Antoni11}.

Differently, the study of the diffractive Drell-Yan reaction performed in \cite{KPST06}
within the light-cone dipole description revealed importance of soft interactions with
the partons spectators, which contributes on the same footing as hard perturbative ones,
and strongly violate QCD factorization.

\begin{figure}[h!]
\centerline{\epsfig{file=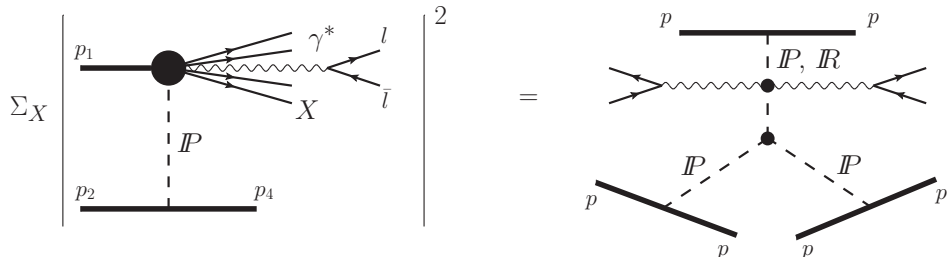,width=13cm}} \caption{The
cross section of the diffractive DY process summed over all
excitation channels at fixed effective mass $M_X$ (left panel)
corresponding to the Mueller graph in Regge picture (right panel).}
\label{fig:Pom}
\end{figure}

Absorptive corrections are normally associated with the soft
interactions between target and projectile, and they play an
important role in diffractive hadron-hadron scattering.
One can derive a Regge behavior of the diffractive cross section
of heavy photon production in terms of the usual light-cone variables,
\begin{eqnarray}
x_{\gamma1} = \frac{p_{\gamma}^+}{p_1^+};\ \ \ \ \
x_{\gamma2} = \frac{p_{\gamma}^-}{p_2^-},
\label{x12}
\end{eqnarray}
so that $x_{\gamma1}x_{\gamma2}=(M^2+k_T^2)/s$ and $x_{\gamma1}-
x_{\gamma2}=x_{\gamma F}$, where $M$, $k_T$ and $x_{\gamma F}$ are
the invariant mass, transverse momentum and Feynman $x_F$ of the
heavy photon (di-lepton). n the limit of small $x_{\gamma1}\to0$ and
large $z_p\equiv p_4^+/p_2^+\to 1$  the diffractive DY cross section
is given by the Mueller graph shown in Fig.~\ref{fig:Pom}. In this
case, the end-point behavior is dictated by the following general
result
\begin{eqnarray}
\frac{d\sigma}{dz_p dx_{\gamma1} dt}\Big|_{t\to0}\propto
\frac{1}{(1-z_p)^{2\alpha_{\Pom}(t)-1}x_{\gamma1}^{\varepsilon}}\,,
\end{eqnarray}
where $\alpha_{\Pom}(t)$ is the Pomeron trajectory corresponding to
the $t$-channel exchange, and $\varepsilon$ is equal to 1 or 1/2 for
the Pomeron $\Pom$ or Reggeon $\Reg$ exchange corresponding to
$\gamma^*$ emission from sea or valence quarks, respectively (see
Fig.~\ref{fig:Pom}).

As an alternative to the factorization based QCD approach, the dipole
description of the QCD diffraction, was presented in Refs.~\cite{KLZ81} (see also
Ref.~\cite{BBGG81}). It is based on the fact that dipoles of different
transverse size $r_{\perp}$ interact with different cross sections
$\sigma(r_{\perp})$, leading to the single inelastic diffractive
scattering with a cross section, which in the forward limit is given
by \cite{KLZ81},
\begin{eqnarray}
\frac{\sigma_{sd}}{dp_{\perp}^2}\Bigg|_{p_{\perp}=0}=
\frac{\langle\sigma^2(r_{\perp})\rangle-\langle\sigma(r_{\perp})\rangle^2}{16\pi},\,
\end{eqnarray}
where $p_{\perp}$ is the transverse momentum of the recoil proton,
$\sigma(r_{\perp})$ is the universal dipole-proton cross section, and operation $\langle...\rangle$ means averaging over
the dipole separation.

The color dipole description of Drell-Yan inclusive process first introduced in
Ref.~\cite{deriv1} (see also Ref.~\cite{BHQ}), treats the production of a heavy di-lepton like photon bremsstrahlung, rather than $\bar qq$ annihilation. Such a difference is a consequence of Lorentz non-invariance of the space-time description of the interaction, which varies with the  reference frame.
Only observables must be Lorentz-invariant.

The dipole approach applied to diffractive Drell-Yan reaction in
Ref.~\cite{KPST06}, led to the QCD factorisation breaking, which
manifests itself in specific features like a significant damping of
the cross section at high $\sqrt{s}$ compared to the inclusive DY
case. This is rather unusual, since a diffractive cross section,
which is proportional to the dipole cross section squared, could be
expected to rise with energy steeper than the total inclusive cross
section, like it occurs in the diffractive DIS process. At the same
time, the ratio of the DDY to DY cross sections was found in
Ref.~\cite{KPST06} to rise with the hard scale, $M^2$. This is also
in variance with diffraction in DIS, which is associated with the
soft interactions \cite{BP97,ourDDIS}.

The absorptive corrections affect differently the diagonal and
off-diagonal terms in the hadronic current \cite{PCAC}, in opposite
directions, leading to an unavoidable breakdown of the QCD
factorisation in processes with off-diagonal contributions only.
Namely, the absorptive corrections enhance the diagonal terms at
larger $\sqrt{s}$, whereas they strongly suppress the off-diagonal
ones. In the diffractive DY process a
new state, the heavy lepton pair, is produced, hence, the whole
process is of entirely off-diagonal nature, whereas in the
diffractive DIS contains both diagonal and off-diagonal
contributions \cite{KPSdiff}. This is the first reason why the QCD
factorisation is broken in the DDY reaction.

The second reason of the QCD factorisation breaking is more specific
and concerns the interplay of soft and hard interactions in the DDY
amplitude. In particular, this leads to the leading twist nature of
the DDY process, whereas DDIS is of the higher twist \cite{KPST06}.
Large and small size projectile fluctuations contribute to the
diffractive DY process at the same footing, which further deepens the
dramatic breakdown of the QCD factorisation in DDY. We will shortly
discuss this issue below when presenting the numerical results.

Quasieikonal model KMR for the so-called ``enhanced'' probability
$\hat{S}_{enh}$ (see e.g. Refs.~\cite{enh-1,enh-2}), frequently used to describe the QCD factorisation breaking in
diffractive processes, is not well justified in higher orders,
whereas the color dipole approach considered here, correctly
includes all diffraction excitations to all orders
\cite{KPSdiff}\footnote{We are thankful to J.~Bartels for pointing
at this issue.}.

In this work, we investigate further the unusual features of the
Drell-Yan diffraction \cite{KPST06} in the framework of the color
dipole approach. We show that the unitarization effects can be
correctly taken into account through eikonalization of the universal
``bare'' elastic dipole-target amplitude. This generalized dipole
approach pretends to take into account the soft absorptive effects
on the same footing with the hard dipole-target scattering. Such
effects are included into the phenomenological partial elastic
dipole amplitude fitted to data. This allows to predict the
diffractive DY cross section completely in terms of the single
parameterization of the dipole cross section known independently
from the soft hadron scattering data.

The paper is organized as follows. Section II contains derivation of
the diffractive Drell-Yan amplitude in the dipole approach. Section
III is devoted to a discussion of the unitarity corrections through
the eikonalization of the elastic ``bare'' dipole-target scattering
amplitude. In Section IV a short overview of different
parameterizations of the elastic dipole-target scattering amplitude
for small and large dipoles is given. The formulae for the single
diffractive Drell-Yan cross section are explicitly derived in
Section V. Discussion of numerical results for the differential
distributions and basic features of the diffractive DY is presented
in Section VI. Finally, Section VII with the summary and conclusions
closes our paper.

\section{Diffractive Drell-Yan amplitude in the dipole approach}

In the forward limit $p_T=0$, the photon radiation from a quark in
inelastic collisions vanishes which was explained at the intuitive level, as well as
 demonstrated by a direct
calculation of Feynman graphs in Ref.~\cite{KST99}. The same is also true
for forward diffraction proceeding via gluon pair exchange
(in the color singlet state) with no momentum transfer between the
projectile quark and the proton target \cite{KPST06}. Disappearance
of both inelastic and diffractive forward photon radiation happens
due to the fact that if the electric charge gets no ``kick'', i.e. is
not accelerated, no photon is radiated, provided that the
radiation time considerably exceeds the duration time of
interaction. This is dictated by the renown Landau-Pomeranchuk
principle \cite{LP}: radiation depends on the strength of the accumulated
kick, rather than on its structure, if the time scale of the kick is
shorter than the radiation time.
\begin{figure}[h!]
\centerline{\epsfig{file=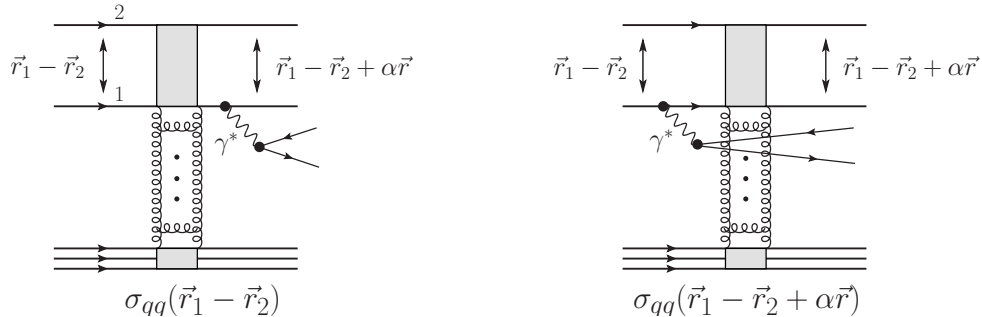,width=13cm}} \caption{Leading
order contribution to the diffractive Drell-Yan in the dipole-target
collision.} \label{fig:gam}
\end{figure}

The non-Abelian case, QCD, is different: a quark can radiate gluons
diffractively in the forward direction. This happens due to
possibility of interaction between the radiated gluon and the
target. Such a process, in particular, is very important for
diffractive heavy flavor production \cite{heavyF}.

Notice that the disappearance of Abelian radiation is only true for the
diffractive scattering of a single quark off the target, and it does
not hold for diffractive hadron-hadron scattering. As was
demonstrated in Ref.~\cite{KPST06}, due to the internal transverse
motion of the valence quarks inside the proton, which corresponds to
finite large transverse separations between them, the forward photon
radiation does not vanish. This means that even at a hard scale the Abelian radiation
is sensitive to the hadron size due
to a dramatic break down of QCD factorization \cite{DL88}. It was
firstly found in Refs.~\cite{Collins93,Collins97} that factorization
for diffractive Drell-Yan reaction fails due to the presence of
spectator partons in the Pomeron. In Ref.~\cite{KPST06} it was
demonstrated that factorization in Drell-Yan diffraction is even
more broken due to presence of spectator partons in the colliding
hadrons.

As usual, we work in the rest frame of the target proton
which remains intact after the collision. The hard part of the
Drell-Yan process is given by the inelastic amplitude of
$\gamma^*$ radiation by a projectile quark (valence or sea) due to its
interaction with the target through a gluon exchange as shown in
Fig.~\ref{fig:gam}. It consists of two terms corresponding to
interaction of two different Fock states with the target -- a bare
quark before the photon emission $|q\rangle$ ($s$-channel diagram),
and a quark accompanied by a Weiz\"acker-Williams photon
$|q\gamma^*\rangle$ ($u$-channel diagram).

Let us consider first heavy photon bremsstrahlung by quark scattered off the proton
target. We imply that the
longitudinal momentum of the projectile quark cannot be changed
significantly at high energies. In the high energy limit the
corresponding $s$ and $u$-channel contributions can be written as
follows \cite{deriv1,RPN02}
\begin{eqnarray}\nonumber
&&{\cal
M}^{\mu}_{s}\simeq-iZ_qe\,\alpha(1-\alpha)\sum_{\sigma}\frac{{\bar
u}_{\sigma_f}(p_f)\gamma^{\mu}u_{\sigma}(p_f+q)}{\alpha^2l_{\perp}^2+\eta^2}\,{\cal
A}_{\sigma\sigma_i}(k_{\perp}),\\&&{\cal M}^{\mu}_{u}\simeq
iZ_qe\,\alpha\sum_{\sigma}\frac{{\bar
u}_{\sigma}(p_i-q)\gamma^{\mu}u_{\sigma_i}(p_i)}
{\alpha^2(\vec{l}_{\perp}+\vec{k}_{\perp})^2+\eta^2}\,{\cal
A}_{\sigma\sigma_f}(k_{\perp}),\quad\eta^2=(1-\alpha)M^2+\alpha^2m_q^2
\end{eqnarray}
where
$\vec{k}_{\perp}=\vec{p}_{f\perp}+\vec{q}_{\perp}-\vec{p}_{i\perp}$
is the transverse momentum of exchanged gluon, and
$\vec{l}_{\perp}=\vec{p}_{f\perp}-(1-\alpha)\vec{q}_{\perp}/\alpha$
is the transverse momentum of the final quark in a frame where
$z$-axis is parallel to the photon momentum, and the amplitude for
scattering of a quark off a nucleon in the rest frame of the nucleon
reads
\begin{eqnarray*}
{\cal A}_{\sigma\sigma_i}(\vec{k}_{\perp})\simeq
2p_i^0\delta_{\sigma\sigma_i}{\cal V}_q(\vec{k}_{\perp}),\qquad
{\cal A}_{\sigma\sigma_f}(\vec{k}_{\perp})\simeq
2p_f^0\delta_{\sigma\sigma_f}{\cal V}_q(\vec{k}_{\perp})
\end{eqnarray*}

Finally, we can switch to impact parameter space performing the
Fourier transformation over $\vec{l}_{\perp}$ and $\vec{k}_{\perp}$
and write down the total amplitude for the photon radiation in
inelastic quark-proton scattering
\begin{eqnarray*}
&&M_q(\vec{b},\vec{r})=-2ip_i^0\,\sqrt{4\pi}\,\frac{\sqrt{1-\alpha}}{\alpha^2}\,
\Psi^{\mu}_{\gamma^*q}(\alpha,\vec{r})\cdot
\Big[V_q(\vec{b})-V_q(\vec{b}+\alpha\vec{r})\Big],\\
&&V_q(\vec{b})=\int\frac{d^2k_{\perp}}{(2\pi)^2}e^{-i\vec{k}_{\perp}\cdot
\vec{b}}{\cal V}_q(\vec{k}_{\perp})\,,
\end{eqnarray*}
where $\alpha\vec{r}$ corresponds to the transverse separation
between initial and final quark induced by the hard photon
radiation, and $\Psi^{\mu}_{\gamma^*q}(\alpha,\vec{r})$ is the
light-cone wave function of the $q\to\gamma^*q$ transition in the
mixed representation defined as follows
\begin{eqnarray*}
\Psi^{\mu}_{\gamma^*q}(\alpha,\vec{r})=Z_q\sqrt{\alpha_{em}}\,
\alpha^3\sqrt{1-\alpha}\int\frac{d^2l_{\perp}}{(2\pi)^2}e^{-i\vec{l}_{\perp}\cdot
\alpha\vec{r}}\,\frac{{\bar
u}_{\sigma_f}(p_f)\gamma^{\mu}u_{\sigma}(p_f+q)}{\alpha^2l_{\perp}^2+\eta^2}
\end{eqnarray*}
The explicit expressions of the LC wave functions products for
radiation of longitudinally ($\lambda=0$) and transversely
($\lambda=\pm1$) polarized photons are \cite{deriv1,BHQ,deriv2}
\begin{eqnarray*}
&&\Psi^{T}_{\gamma^*q}(\alpha,\vec{\rho}_1)
\Psi^{T*}_{\gamma^*q}(\alpha,\vec{\rho}_2)= \sum_{\lambda=\pm
1}\frac{1}{2}\sum_{\sigma_f\sigma_i}
\epsilon^*_\mu(\lambda)\Psi^{\mu}_{\gamma^*q}(\alpha,\vec{\rho}_1)
\epsilon_\nu(\lambda)\Psi^{\nu*}_{\gamma^*q}(\alpha,\vec{\rho}_2)\\
\nonumber &&\qquad= Z_q^2\frac{\alpha_{em}}{2\pi^2}\Bigg\{
     m_q^2 \alpha^4 {\rm K}_0\left(\eta \rho_1\right)
     {\rm K}_0\left(\eta \rho_2\right)+ \left[1+\left(1-\alpha\right)^2\right]\eta^2
   \frac{\vec{\rho}_1\cdot\vec{\rho}_2}{\rho_1\rho_2}
     {\rm K}_1\left(\eta \rho_1\right)
     {\rm K}_1\left(\eta \rho_2\right)\Bigg\},\\ \nonumber
&&\Psi^{L}_{\gamma^*q}(\alpha,\vec{\rho}_1) \Psi^{L*}_{\gamma^*
q}(\alpha,\vec{\rho}_2)= \frac{1}{2}\sum_{\sigma_f\sigma_i}
\epsilon^*_\mu(\lambda=0)\Psi^{\mu}_{\gamma^*
q}(\alpha,\vec{\rho}_1)
\epsilon_\nu(\lambda=0)\Psi^{\nu*}_{\gamma^* q}(\alpha,\vec{\rho}_2)\\
&&\qquad= Z_q^2\frac{\alpha_{em}}{\pi^2}M^2 \left(1-\alpha\right)^2
  {\rm K}_0\left(\eta \rho_1\right)
     {\rm K}_0\left(\eta \rho_2\right).
\end{eqnarray*}

Now let us turn to elastic dipole scattering as depicted in
Fig.~\ref{fig:gam}. It corresponds to forward scattering at small
momentum transfers in the $t$-channel. Generally speaking,
$\sqrt{-t}\to \Lambda_{QCD}$ corresponds to the physical forward
scattering limit since transverse momentum of a proton in the final
state cannot be resolved to a better accuracy than its inverse size.

In the leading order the elastic scattering amplitude is given by
one-loop diagram with two $t$-channel gluon exchanges. Due to
on-shell intermediate spectators, corresponding four-dimensional
loop integral can be reduced to two-dimensional one over the
transverse momentum of one of the gluons
\begin{eqnarray}
2i\mathrm{Im}\,F_{el}(\vec{\delta}_{\perp})=\int\frac{d^2k_{\perp}}{(2\pi)^2}\,
A(\vec{k}_{\perp})A(\vec{\delta}_{\perp}-\vec{k}_{\perp}),\qquad
\vec{\delta}_{\perp}\ll |\vec{k}_{\perp}|
\end{eqnarray}
where $A$ represents (inelastic) amplitude for one $t$-channel gluon
exchange, and the last strong inequality guarantees that the proton
target survives the scattering, hence, the elastic nature of the
process. Then the convolution theorem of Fourier analysis leads to
the optical theorem
\begin{eqnarray}
\mathrm{Im}\,F_{el}(\vec{\delta})=\int
d^2b\,e^{-i\vec{\delta}_{\perp}\cdot\vec{b}}\,\mathrm{Im}\,f_{el}(\vec{b}),\qquad
2i\mathrm{Im}\,f_{el}(\vec{b})=|\tilde{A}(\vec{b})|^2
\label{theorem}
\end{eqnarray}
which will be used below for eikonalization of multiple elastic
amplitudes.

Repeating calculations in this case we arrive at the $\bar qq$ dipole
scattering amplitudes for $s$ and $u$-channel photon emission,
respectively,
\begin{eqnarray*}
M^{(1)s}_{\bar qq}(\vec{b},\vec{r}_p,\vec{r},\alpha)&=&-2ip_i^0\,\sqrt{4\pi}\,\frac{\sqrt{1-\alpha}}{\alpha^2}\,
\Psi^{\mu}_{\gamma^*q}(\alpha,\vec{r})\\
&&\times\frac{1}{N_c}\sum_X\sum_{c_fc_i}\left(\big|V_q(\vec{b})-V_q(\vec{b}+\vec{r_p})\big|^2-\big|V_q(\vec{b}+\vec{r_p})\big|^2\right),\\
M^{(1)u}_{\bar qq}(\vec{b},\vec{r}_p,\vec{r},\alpha)&=&2ip_i^0\,\sqrt{4\pi}\,\frac{\sqrt{1-\alpha}}{\alpha^2}\,
\Psi^{\mu}_{\gamma^*q}(\alpha,\vec{r})\\
&&\times\frac{1}{N_c}\sum_X\sum_{c_fc_i}
\left(\big|V_q(\vec{b})-V_q(\vec{b}+\vec{r_p}+\alpha\vec{r})\big|^2-\big|V_q(\vec{b}+\vec{r_p}+\alpha\vec{r})\big|^2\right),
\end{eqnarray*}
where the last terms subtract the contributions from diagrams
corresponding to the situation when none of the gluons couple to the
same quark line with the hard photon. Then, implied the fact that
all fields disappear at infinite separations, i.e.
$V_q(\vec{b})\to0$ when $|\vec{b}|\to\infty$, we have due to
antisymmetry of the integrand
\begin{eqnarray}\label{extra-terms}
\int d^2b\,e^{-i\vec{\delta}_{\perp}\cdot\vec{b}}
\Big[\big|V_q(\vec{b}+\vec{r_p})\big|^{2n}-\big|V_q(\vec{b}+\vec{r_p}+\alpha\vec{r})\big|^{2n}\Big]\to0,\quad
n\geq 1\,,\quad |\vec{\delta}_{\perp}|\to0\,,
\end{eqnarray}
such that these terms do not contribute to the final result. Using
the optical theorem for the elastic amplitude
\begin{eqnarray*}
2i\,\mathrm{Im}\,
f_{el}(\vec{b},\vec{r}_p)=\frac{i}{N_c}\sum_X\sum_{c_fc_i}\,\big|V_q(\vec{b})-V_q(\vec{b}+\vec{r_p})\big|^2,
\end{eqnarray*}
we can finally write
\begin{eqnarray}
M^{(1)}_{\bar qq}(\vec{b},\vec{r}_p,\vec{r},\alpha)=-2ip_i^0\,\sqrt{4\pi}\,\frac{\sqrt{1-\alpha}}{\alpha^2}\,
\Psi^{\mu}_{\gamma^*q}(\alpha,\vec{r})\left[2\mathrm{Im}\,
f_{el}(\vec{b},\vec{r_p})-2\mathrm{Im}\,
f_{el}(\vec{b},\vec{r}_p+\alpha\vec{r})\right]\label{amp-LO}
\end{eqnarray}
i.e. the amplitude of the diffractive radiation is proportional to
the difference between elastic amplitudes of the two Fock
components, with and without the photon radiation. When a quark
fluctuates into the upper Fock quark-photon state with the
transverse separation $\vec{r}$, the final quark gets a transverse
shift $\Delta\vec{r}=\alpha\vec{r}$. Then the quark dipoles with
different sizes in the $|2q\rangle$ and $|2q\gamma^*\rangle$
components interact differently, and their difference corresponds to
the diffractive Drell-Yan process amplitude (\ref{amp-LO}).

\section{Breakdown and restoration of unitarity}

The elastic hadron-hadron scattering, which is observed in
an experiment,  is a complicated process, which can be composed to many {\it elementary}
(``bare'') elastic scatterings which are, in fact, the shadows of
many inelastic interactions to be resummed to all orders. Such a
resummation of the elementary scatterings, leads to {\it
unitarization} of the elastic amplitude.

By definition, an eigenstate of interaction cannot be diffractively excited, so can experience
only multiple elastic interactions. Correspondingly, its interaction cross section can be eikonalized, and this is not an approximation (like Glauber model for hadronic interaction),
but is the exact result. In high-energy QCD the set of eigenstates are identified with
the Fock states which can be treated as color dipoles. In the Drell-Yan reaction the lowest Fock state is an effective $|\bar qq\rangle$ dipole \cite{deriv1}. Higher Fock states, like $|\bar qqg\rangle$, etc. also contribute and their amplitudes should also be eikonalized.
The approximation used here is to neglect those corrections. This is is justified for not very small fraction
$x_{\gamma1}$ and scale $M^2$, where valence/sea quarks are dominated
and the gluon contribution is rather small.

In terms of the Regge theory, the elementary elastic
dipole-target scattering corresponds to an exchange of the ``bare''
Pomeron. Assuming that this bare Pomeron is a Regge pole with the intercept above one (to justify the observed rising
energy dependence of the cross section) one breaks down unitarity (Froissart bound) in the high energy limit. Eikonalization of this "bare" amplitude restores unitarity. The mentioned above higher Fock states correspond in Regge description to so called enhanced Regge graphs. Summing up all such graphs one arrives to the "effective" Pomeron, called Froissaron \cite{dubovikov}, which satisfies the unitarity restrictions in both $s$-channel and $t$-channels. An additional justification to the mentioned above approximation neglecting higher Fock states, or enhanced graphs, comes from the observed smallness of the triple-Pomeron coupling.

\begin{figure}[h!]
\centerline{\epsfig{file=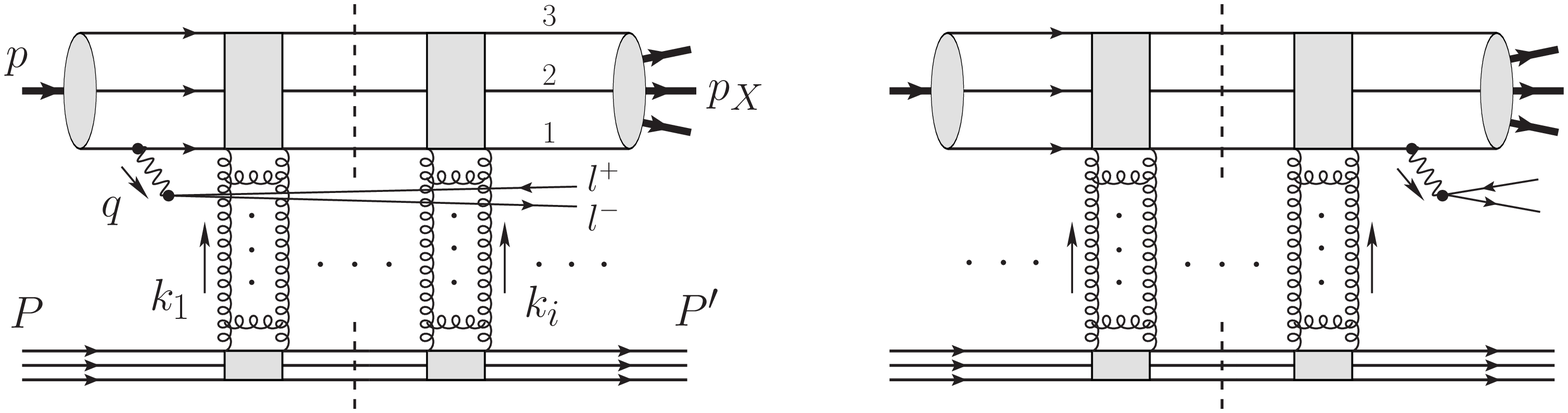,width=14cm}} \caption{Total
eikonalized amplitude of the diffractive Drell-Yan process, which
effectively includes the gap survival effects.} \label{fig:eik}
\end{figure}

Having all this in mind, we can now easily generalize the
expression (\ref{amp-LO}) for the case of $n$ consequent
``bare'' elastic scatterings of spectators as shown in
Fig.~\ref{fig:eik}. The amplitude in this case is given by
\begin{eqnarray}
M^{(n)}_{\bar qq}(\vec{b},\vec{r}_p,\vec{r},\alpha)\sim
\frac{(-1)^n[2\mathrm{Im}\,
f_{el}(\vec{b},\vec{r_p})]^n}{n!}-\frac{(-1)^n[2\mathrm{Im}\,
f_{el}(\vec{b},\vec{r_p}+\alpha\vec{r})]^n}{n!} \label{amp-n}
\end{eqnarray}
The sign factor $(-1)^n$ is related to the phase of the bare Pomeron amplitude, which is nearly $\pi/2$ due to the absorptive origin of the Pomeron, which is generated  in the elastic amplitude by inelastic collisions through the unitarity relation. This can be also seen as a consequence of QCD as a non-Abelian theory \cite{KPSdiff}. Indeed, if it were an Abelian theory, the Born graph for the elastic amplitude would be one gluon exchange, i.e. the amplitude would
be real. Since, however, QCD is non-abelian, the minimal number of exchanged gluons is two, i.e. the amplitude is imaginary.

In (\ref{amp-n}) we followed neglected the
contributions with the photon radiated by the quark between subsequent
elastic scatterings, in accordance with the Landau-Pomeranchuk
principle mentioned above, which is at work if the radiation with a long coherence length,
\begin{eqnarray}
l_c=\frac{1}{2x_{\gamma2}m_N}\gg L,
\label{tc}
\end{eqnarray}
where $x_{\gamma2}$ is defined in (\ref{x12}) and $L$ is the longitudinal distance covered by the interaction. In this case the radiation between the multiple interactions is suppressed by interferences (compare with gluon radiation in \cite{bh-eloss}), and the radiation spectrum
depends only on the total momentum transfer, rather than on its multiple structure.
Therefore multiple interactions do not lift the ban on a forward diffractive Abelian radiation,
in spite of nonzero momenta transferred in each of the multiple scatterings.

Resummation of
amplitudes (\ref{amp-n}) to all rescattering orders
results in the total amplitude, which
has the following general form
\begin{eqnarray}
M_{\bar qq}^{\mu}(\vec{b},\vec{r}_p,\vec{r},\alpha)=2ip_i^0\,\sqrt{4\pi}\,\frac{\sqrt{1-\alpha}}{\alpha^2}\,
\Psi^{\mu}_{\gamma^*q}(\alpha,\vec{r})\left[e^{-2\mathrm{Im}\,
f_{el}(\vec{b},\vec{r_p})}-e^{-2\mathrm{Im}\,
f_{el}(\vec{b},\vec{r}_p+\alpha\vec{r})}\right]\label{amp-tot}
\end{eqnarray}
Note, in this derivation we implied that relation
(\ref{extra-terms}) holds in each particular order $n$, thus the
subtraction terms do not contribute to the final expression
(\ref{amp-tot}).

Let us now write the total hadronic amplitude of the diffractive
Drell-Yan process as \cite{KPST06}
\begin{eqnarray}
A_{if}=A^{(1)}_{if}+A^{(2)}_{if}+A^{(3)}_{if}\,,
\end{eqnarray}
where each term corresponds to $\gamma^*$ radiation by one of the
valence quarks, in particular,
\begin{eqnarray}
A^{(1)}_{if}(x_{\gamma1},\vec{q}_{\perp},\lambda_{\gamma})&=&\frac{i}{4}\,\alpha^2\int
d^2r_1d^2r_2d^2r_3d^2rd^2bdx_{q_1}dx_{q_2}dx_{q_3}\nonumber\\
&\times&\Psi_{i}(\vec{r}_1,
\vec{r}_2,\vec{r}_3;x_{q_1},x_{q_2},x_{q_3})
\Psi_{f}^*(\vec{r}_1+\alpha
\vec{r},\vec{r}_2,\vec{r}_3;x_{q_1}-x_{\gamma1},x_{q_2},x_{q_3})\nonumber\\
&\times&\Big[M^{\lambda_{\gamma}}_{\bar qq}(\vec{b},\vec{r}_1-\vec{r}_2,\vec{r},\alpha)+
M^{\lambda_{\gamma}}_{\bar qq}(\vec{b},\vec{r}_1-\vec{r}_3,\vec{r},\alpha)\Big]
e^{i\vec{l}_{\perp}\cdot\alpha\vec{r}}e^{i\vec{\delta}_{\perp}\cdot\vec{b}}
\label{amp-DDY}
\end{eqnarray}
Here, $\lambda_{\gamma}=L,T$;
$\vec{l}_{\perp}=\vec{\delta}_{\perp}-\vec{q}_{\perp}/\alpha$
($z$-axis is directed along initial proton momentum); the hard
photon with virtuality $q^2=M^2\gg m_p^2$, transverse
$\vec{q}_{\perp}$ and fractional longitudinal $x_{\gamma1}$ momenta
is emitted from the first valence quark with impact parameter
$\vec{r}_1$ (see Fig.~\ref{fig:eik}), other two valence quarks in
the proton have impact parameters $\vec{r}_2$ and $\vec{r}_3$,
respectively; $\vec{r}$ is transverse separation between the photon
and the radiating quark; $\alpha=x_{\gamma1}/x_{q_1}$ is the
fraction of longitudinal momenta taken away by the photon from the
radiating quark; $M^{L,T}_{\bar qq}$ are the Fourier-transformed
amplitudes for the elastic quark dipole scattering off the proton
target accompanied by the hard $L,T$-polarized photon emission
calculated above in Eq.~(\ref{amp-tot}); $\Psi_{i,f}$ are the
light-cone wave functions of the $3q$ systems in the initial and
final state, respectively. In Eq.~(\ref{amp-DDY}) we implicitly
assumed that exchanges $t$-channel gluons all together  take a
negligibly small longitudinal momentum compared to the collisions
energy $\sqrt{s}$ and, hence, corrections to quark momenta due to
gluon couplings are neglected in the wave functions.

Eikonalization of the universal elastic dipole-target scattering
amplitude performed in Eq.~(\ref{amp-tot}) incorporates
all soft and hard interactions between $\bar qq$ dipole and the target on
the same footing as the parameterization for the amplitude on a free proton,
$f_{el}(\vec{b},\vec{r})$, fitted to data on DIS and soft hadron scattering.
This provides an alternative to the conventional treatment of
the gap survival effects in terms of the single suppression factor
in the cross section $\sigma_{DDY}=K\cdot\sigma_{DDY}^{bare}$, the
so-called gap survival factor, estimated in the eikonal approximation \cite{KPS06}
\begin{eqnarray}
K=1-\frac{1}{\pi}\frac{\sigma_{tot}^{pp}(s)}{B_{sd}^{DY}(s)+2B_{el}^{pp}(s)}+
\frac{1}{(4\pi)^2}\frac{[\sigma_{tot}^{pp}(s)]^2}
{B_{el}^{pp}(s)[B_{sd}^{DY}(s)+B_{el}^{pp}(s)]} \label{gapS}
\end{eqnarray}
where the energy-dependent elastic slope is
$B_{el}^{pp}(s)=B_{el}^0+2\alpha_{\Pom}'\ln(s/s_0)$ with
$B_{el}^0=7.5\,\GeV^{-2},\,s_0=1\,\GeV^2$. The slope of
single-diffractive DY cross section can be estimated as,
$B_{sd}^{DY}\simeq \langle
r_{ch}^2\rangle/3+2\alpha_{\Pom}'\ln(s/s_0)$, where the proton mean
charge radius squared $\langle r_{ch}^2\rangle=0.8$ fm$^2$. More
elaborated models for the gap survival factor incorporating a part
of the Gribov corrections (see e.g. Refs.~\cite{gotsman,kaidalov})
predict similar suppression factors, and one can easily replace the
$K$-factor (\ref{gapS}) by a preferable one.

In order to demonstrate that the eikonalization procedure
(\ref{amp-tot}) correctly takes into account soft gap survival
effects, we checked that the ratio between diffractive Drell-Yan
cross sections with eikonalized (\ref{amp-tot}) and non-eikonalized
(\ref{amp-LO}) diffractive amplitude leads to a suppression factor
which is very close numerically to the standard gap survival factor
$K$ defined independently from Eq.~(\ref{gapS}),
\begin{eqnarray}
\frac{\sigma_{DDY}^{eik}}{\sigma_{DDY}^{noneik}}\simeq K
\end{eqnarray}
and such a relation between them holds at different energies. And
this ratio does not depend on $x_{\gamma1}$ and $M^2$. This confirms
our statement we have made in the beginning of this Section: the
eikonalization of the elastic dipole-target amplitude correctly
incorporates the unitarity corrections.

As was already been said above, the parameterizations for the real
elastic dipole-target scattering fitted to experimental data must
already effectively contain the absorptive and all-order QCD
corrections, so it should not be eikonalized when used in explicit
calculations, and no $K$-factors are to be applied to the cross
section in this case.

Let us now shortly discuss different parameterizations for the
elastic dipole-target scattering amplitude $f_{el}(\vec{b},\vec{r})$
corresponding to the scattering of small ($r\ll r_p$) and large
($r\sim r_p$) dipoles off the proton target, known from the data
fits.

\section{Elastic dipole-target scattering amplitude}

It is well-known that small $x$ (large $Q^2$) regime corresponding
to a scattering of small dipoles with $r\to0$ is well described by
the popular Golec-Biernat-Wuestoff (GBW) parametrization of the
dipole cross section \cite{GBWdip}. The elastic $b$-dependent
amplitude in this case is $x$-dependent and has a form
\cite{GBW-par} (see also Ref.~\cite{KST-GBW-eqs})
\begin{eqnarray}\nonumber
&&\mathrm{Im}f_{el}^{\rm{GBW}}(\vec{b},\vec{r},x,x_q)=\frac{\sigma_0}{8\pi{\cal
B}(x)}\Bigg\{\exp\Bigg[-\frac{[\vec{b}+\vec{r}(1-x_q)]^2}{2{\cal
B}(x)}\Bigg]+\exp\Bigg[-\frac{[\vec{b}+\vec{r}x_q]^2}{2{\cal
B}(x)}\Bigg]\\&&-\,2\exp\Bigg[-\frac{r^2}{R_0^2(x)}-\frac{[\vec{b}+\vec{r}(1/2-x_q)]^2}{2{\cal
B}(x)}\Bigg]\Bigg\},\qquad {\cal
B}(x)=R_N^2(x)+R_0^2(x)/8\,,\label{GBW}
\end{eqnarray}
where parameters are fitted to DIS data at small $x$ \cite{GBWdip}
$\sigma_0=23.03$ mb, $R_0(x)=0.4\,\mathrm{fm}\times(x/x_0)^{0.144}$
with $x_0=3.04\times10^{-4}$, $x\sim Q^2/s$ is the Bjorken variable,
$x_q$ is the quark longitudinal quark fraction in the dipole defined
in Eq.~(\ref{SF}), $R_N^2(x)$ in the limit $r\to0$ can be defined
through the slope of elastic electroproduction of $\rho$-mesons
measured at HERA as
\[R_N^2(x)=B_{\gamma^*p\to \rho p}(x,Q^2\gg 1\,\GeV^2)-R_0^2(x)/4,\quad
B_{\gamma^*p\to \rho p}(x,Q^2\gg 1\GeV^2)\simeq 5\,\GeV^{-2}\]\,.
Amplitude (\ref{GBW}) correctly reproduces the dipole
cross section \cite{GBWdip}
\[2\int d^2b\,\mathrm{Im}f_{el}(\vec{r},\vec{b},x,x_q)=\sigma_{\bar qq}(r,x)\,,\]
but contains more information, because it is sensitive to the color
dipole orientation within the phenomenological saturation model,
which includes contributions from higher order perturbative
corrections as well as non-perturbative effects contained in DIS
data.

Notice that the simple GBW parameterization (\ref{GBW}) has some
restrictions. In particular, the non-integrated gluon distribution
 exhibits no power-law tails in
momentum space in contradiction with QCD. Moreover, it does not
match the DGLAP evolution at large values of $Q^2$. Therefore, one
should be cautious applying this model at very high transverse
momenta accessible at the energies of LHC \cite{KST-GBW-eqs}.

For soft scattering (moderate and small $Q^2$) corresponding to
large dipoles $r\sim R_0(x)$, the c.m. energy squared $s$, rather
than Bjorken $x$, is the proper variable. So, for the soft processes
one can switch from $x$- to $s$-dependence \cite{KST99}, keeping the
same functional form of the dipole amplitude (\ref{GBW}) and
adjusting the parameters to observables in soft reactions as
\cite{kpss,KST-par,KST-GBW-eqs}
\begin{eqnarray}\label{KST}
\mathrm{Im}f^{\mathrm{GBW}}_{el}(\vec{b},\vec{r},x,x_q)\quad&\rightarrow&\quad
\mathrm{Im}f^{\mathrm{KST}}_{el}(\vec{b},\vec{r},s,x_q),\\
R_0(x)\quad&\rightarrow&\quad
R_0(s)=0.88\,\mathrm{fm}\,(s_0/s)^{0.14},\nonumber\\
R_N^2(x)\quad&\rightarrow&\quad R_N^2(s)=B_{el}^{\pi
p}(s)-\frac14R_0^2(s)-\frac13\langle r_{ch}^2 \rangle_{\pi},\nonumber\\
\sigma_0\quad&\rightarrow&\quad \sigma_0(s)=\sigma_{tot}^{\pi
p}(s)\Big(1+\frac{3R_0^2(s)}{8\langle r_{ch}^2
\rangle_{\pi}}\Big),\nonumber
\end{eqnarray}
where the pion-proton total cross section is parameterized as
\cite{barnett} $\sigma_{tot}^{\pi p}(s)=23.6(s/s_0)^{0.08}$ mb,
$s_0=1000\,\GeV^2$, the mean pion radius squared is \cite{amendolia}
$\langle r_{ch}^2 \rangle_{\pi}=0.44$ fm$^2$, and the Regge
parametrization of the elastic slope $B_{el}^{\pi
p}(s)=B_0+2\alpha'_{\Pom}\ln(s/\mu^2)$, with $B_0=6\,\GeV^{-2}$,
$\alpha'_{\Pom}=0.25\,\GeV^{-2}$, and $\mu^2=1\,\GeV^2$ can be used.
We shall refer to
$\mathrm{Im}f^{\mathrm{KST}}_{\bar qq}(\vec{b},\vec{r},s,x_q)$ below as
the Kopeliovich-Sch\"afer-Tarasov (KST) parametrization of the elastic
dipole-proton amplitude \cite{KST-par}.

Two models for the elastic quark dipole-target scattering amplitude
$f_{el}(\vec{b},\vec{r})$, given above by Eqs.~(\ref{GBW}) and
(\ref{KST}), are valid for small $r\ll r_p$ and large
$r\sim r_p$ dipole ($r_p$ is the mean proton size), respectively.
Since the diffractive DY cross section is primarily sensitive to the
large transverse  separations $\sim r_p$ between target and projectile
implied by the forward limit $t\to 0$, then the KST parameterization
for the dipole cross section should be used, at least, in the
leading order calculation. For completeness, we will compare the DDY
cross sections calculated with GBW and KST parameterizations below
when discussing the numerical results.

\section{Single diffractive Drell-Yan cross section}

The differential cross section for the single diffractive di-lepton
production in the target rest frame reads
\begin{eqnarray}\nonumber
d^{\,8}\sigma^{sd}_{\lambda_{\gamma}}(pp\to
pl^+l^-X)&=&\sum_f\sum_{n=1}^3
|A^{(n)}_{if}(x_{\gamma1},\vec{q}_{\perp},\lambda_{\gamma})|^2\frac{d\alpha}{\alpha(1-\alpha)}\frac{
d^2q_{\perp}d^2\delta_{\perp}}{(2\pi)^5\,8(p_{i,n}^0)^2}
\\
&\times&\alpha_{em}\epsilon_{\mu}(\lambda_{\gamma})\epsilon^*_{\nu}(\lambda_{\gamma})L^{\mu\nu}
\frac{dM^2d\Omega}{16\pi^2M^4},\quad \lambda_{\gamma}=L,T
\label{cross-sec}
\end{eqnarray}
where prefactors provide averaging over colors and helicities of
exchanged $t$-channel gluons, $p_{i,n}^0$ is the energy of the
radiating $n$th quark in the initial state, $n=1,\,...,\,3$;
$\alpha_{em}=e^2/(4\pi)=1/137$ is the electromagnetic coupling
constant. The second line in Eq.(\ref{cross-sec}) describes decay of
$\gamma^*$ into the leptonic pair $l^+l^-$ into solid angle
$d\Omega=d\phi d\cos\theta$, and the standard leptonic tensor is
given by
\begin{eqnarray}\nonumber
L^{\mu\nu}=4(p_{l^+}^{\mu}p_{l^-}^{\nu}+p_{l^+}^{\nu}p_{l^-}^{\mu}-g^{\mu\nu}(p_{l^+}\cdot
p_{l^-}))\,.
\end{eqnarray}
At the moment, we are not interested in lepton polarizations and
their angular distributions, so for the sake of simplicity we integrate out the
cross section (\ref{cross-sec}) over the solid angle of the lepton
pair. We keep in the cross section only diagonal in
the photon polarization $\lambda_{\gamma}=L,T$ terms (non-diagonal ones
drop out after integration over leptonic azimuthal angle $\phi$).
Integrating the diffractive differential DY cross section over the
photon transverse momentum $\vec{q}_{\perp}$ we get
\begin{eqnarray}\label{DDY-cs}
\frac{d^4\sigma_{L,T}(pp\to
pl^+l^-X)}{d\ln\alpha\,dM^2\,d^2\delta_{\perp}}=
\frac{\alpha_{em}}{3\pi M^2}\,\frac{d^3\sigma_{L,T}(pp\to
p\gamma^*X)}{d\ln\alpha\,d^2\delta_{\perp}}\,.
\end{eqnarray}
Then applying the completeness relation
\begin{eqnarray}\nonumber
&&\sum_f\Psi_f(\vec{r}_1+\alpha
\vec{r},\vec{r}_2,\vec{r}_3;x_{q_1},x_{q_2},x_{q_3})
\Psi^*_f(\vec{r}\,'_1+\alpha
\vec{r}\,',\vec{r}\,'_2,\vec{r}\,'_3;x'_{q_1},x'_{q_2},x'_{q_3})\\
&&\phantom{.......}=\,
\delta(\vec{r}_1-\vec{r}\,'_1+\alpha(\vec{r}-\vec{r}\,'))\delta(\vec{r}_2-\vec{r}\,'_2)
\delta(\vec{r}_3-\vec{r}\,'_3)\prod_{j=1}^3\delta(x_{q_j}-x'_{q_j})
\end{eqnarray}
we get the diffractive $\gamma^*$ production cross section in the
following differential form
\begin{eqnarray}
&&\frac{d^3\sigma_{\lambda_{\gamma}}(pp\to
p\gamma^*X)}{d\ln\alpha\,d^2\delta_{\perp}}=\frac{\sum_q
Z_q^2}{64\pi^2}\int
d^2r_1d^2r_2d^2r_3d^2r\,d^2bd^2b'\,dx_{q_1}dx_{q_2}dx_{q_3}\nonumber\\
&&\qquad\qquad\times\,|\tilde{\Psi}^{\lambda_{\gamma}}_{\gamma^*q}(\alpha,\vec{r})|^2|\Psi_{i}(\vec{r}_1,
\vec{r}_2,\vec{r}_3;x_{q_1},x_{q_2},x_{q_3})|^2\nonumber\\
&&\qquad\qquad\times\,\Delta(\vec{r}_1,
\vec{r}_2,\vec{r}_3;\vec{b};\vec{r},\alpha)\Delta(\vec{r}_1,
\vec{r}_2,\vec{r}_3;\vec{b}\,';\vec{r},\alpha)\,
e^{i\vec{\delta}_{\perp}\cdot(\vec{b}-\vec{b}\,')} \label{eik-tot}
\end{eqnarray}
where $\tilde{\Psi}_{\gamma^*q}=\Psi_{\gamma^*q}/Z_q$, and
\begin{eqnarray}\nonumber
\Delta&=&-2\mathrm{Im}\,
f^{\mathrm{KST}}_{el}(\vec{b},\vec{r}_1-\vec{r}_2)+2\mathrm{Im}\,
f^{\mathrm{KST}}_{el}(\vec{b},\vec{r}_1-\vec{r}_2+\alpha\vec{r})\\&&-2\mathrm{Im}\,
f^{\mathrm{KST}}_{el}(\vec{b},\vec{r}_1-\vec{r}_3)+2\mathrm{Im}\,
f^{\mathrm{KST}}_{el}(\vec{b},\vec{r}_1-\vec{r}_3+\alpha\vec{r})\,,
\label{eik-el}
\end{eqnarray}
where the KST parameterization (\ref{KST}) fitted to the soft data
and hence valid at $|\vec{r}_i-\vec{r}_j|\sim \vec{b},\,i\not=j$ is
used. Finally, going over to the forward limit
$\vec{\delta}_{\perp}=0$ we get
\begin{eqnarray}
&&\frac{d^3\sigma_{\lambda_{\gamma}}(pp\to
p\gamma^*X)}{d\ln\alpha\,d\delta_{\perp}^2}\Big|_{\delta_{\perp}=0}=\frac{\sum_q
Z_q^2}{64\pi}\int
d^2r_1d^2r_2d^2r_3d^2r\,dx_{q_1}dx_{q_2}dx_{q_3}\nonumber\\
&&\qquad\times\,|\tilde{\Psi}^{\lambda_{\gamma}}_{\gamma^*q}
(\alpha,\vec{r})|^2|\Psi_{i}(\vec{r}_1,
\vec{r}_2,\vec{r}_3;x_{q_1},x_{q_2},x_{q_3})|^2\,\left[\int d^2b\,
\Delta(\vec{r}_1,
\vec{r}_2,\vec{r}_3;\vec{b};\vec{r},\alpha)\right]^2
\end{eqnarray}
We see that normalization of the cross section agrees with the
original result of Ref.~\cite{KPST06}. The total diffractive cross
section is then given by
\begin{eqnarray}
\frac{d\sigma(pp\to
p\gamma^*X)}{d\ln\alpha}=\frac{1}{B_{sd}^{DY}(s)}\frac{d^3\sigma(pp\to
p\gamma^*X)}{d\ln\alpha\,d\delta_{\perp}^2}\Big|_{\delta_{\perp}=0}
\end{eqnarray}
where $B_{sd}^{DY}(s)$ is the diffractive slope similar to the one
measured in diffractive DIS.

The next step is to introduce the proton wave function assuming the
Gaussian shape for the quark distributions in the proton as
\begin{eqnarray}\nonumber
|\Psi_i(\vec{r}_1,
\vec{r}_2,\vec{r}_3;x_{q_1},x_{q_2},x_{q_3})|^2&=&\frac{2+a/b}{\pi^2ab}
\exp\Big[-\frac{r_1^2}{a}-\frac{r_2^2+r_3^2}{b}\Big]\rho(x_{q_1},x_{q_2},x_{q_3})\\
&\times&\delta(\vec{r}_1+\vec{r}_2+\vec{r}_3)\delta(1-x_{q_1}-x_{q_2}-x_{q_3})
\label{psi}
\end{eqnarray}
where $a=\langle r_{\bar qq}^2 \rangle$ and $b=\langle R_q^2 \rangle$ are
the diquark mean radius squared and the quark mean distance from the
diquark squared, respectively.
In this work, we will use the simplest case of symmetric valence
quarks distribution assuming that $r_{\bar qq}=R_q=0.85$ fm.

Then valence quark distribution in the proton is given by
\begin{eqnarray}\nonumber
\int
dx_{q_2}dx_{q_3}\rho(x_{q_1},x_{q_2},x_{q_3})=\rho_{q_1}(x_{q_1})\,.
\end{eqnarray}
where we integrated out the longitudinal fractions of the diquark in
the proton. Generalization of the three-body proton wave function
(\ref{psi}) including different quark and antiquark flavors leads to
the proton structure function \cite{KRT00}
\begin{eqnarray}\label{SF}
\sum_q Z_q^2[\rho_q(x_q)+\rho_{\bar
q}(x_q)]=\frac{1}{x_q}F_2(x_q),\qquad
x_q=\frac{x_{\gamma1}}{\alpha}\,.
\end{eqnarray}
In the numerical analysis below, in order to estimate the
theoretical uncertainty in the diffractive DY process we will use a
few different parameterizations for the proton structure function
$F_2$, widely used in the literature.

\section{Numerical results}

\begin{figure}[!h]
\begin{minipage}{0.49\textwidth}
 \centerline{\includegraphics[width=1.0\textwidth]{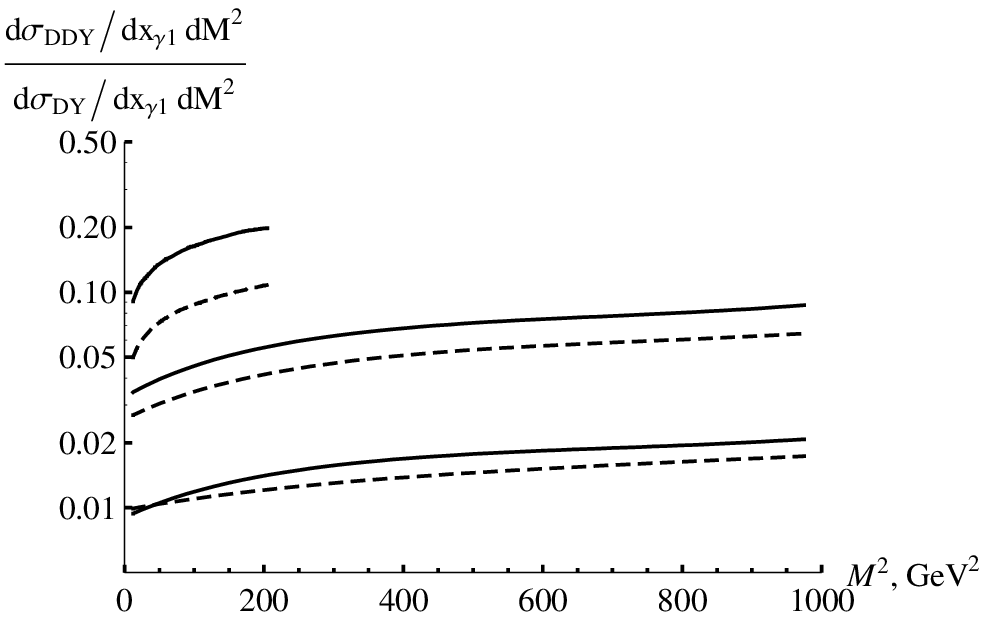}}
\end{minipage}
\hspace{0.5cm}
\begin{minipage}{0.46\textwidth}
 \centerline{\includegraphics[width=1.0\textwidth]{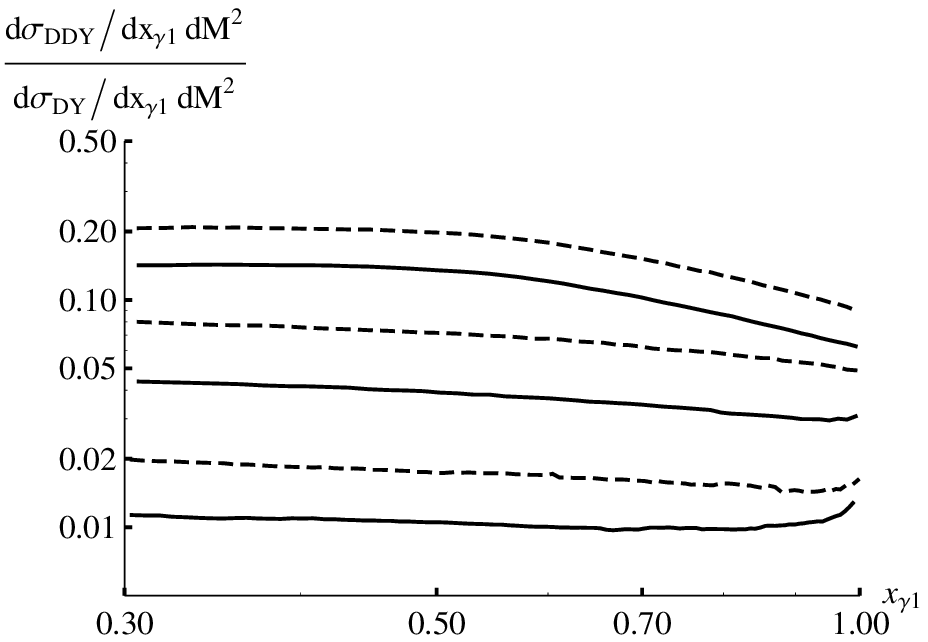}}
\end{minipage}
   \caption{
\small The ratio of the diffractive to inclusive Drell-Yan cross
sections as function of the lepton-pair invariant mass squared
$M^2$ (left panel) and photon fraction $x_{\gamma1}$ (right panel) at
different energies. In the left panel, the curves are given for
fixed $x_{\gamma1}=0.5$ (solid lines) and $x_{\gamma1}=0.9$ (dashed
lines). In the right panel, the curves are given for fixed
$M^2=50\,\GeV^2$ (solid lines) and $M^2=500\,\GeV^2$ for
$\sqrt{s}=14$ TeV and 500 GeV, and $M^2=200\,\GeV^2$ at
$\sqrt{s}=40$ GeV (dashed lines). The pairs of solid/dashed curves
in the both panels correspond to $\sqrt{s}=40\,\GeV$, $500$ GeV and
$14$ TeV from top to bottom, respectively.}
 \label{fig:DDYvsDY}
\end{figure}

Let us now turn to discussion of the numerical results. We start
from the comparison of the differential cross sections for single
diffractive and inclusive Drell-Yan processes. In
Fig.~\ref{fig:DDYvsDY} the ratio of the diffractive to inclusive DY
cross sections is plotted as a function of di-lepton invariant mass
squared $M^2$ (left panel) and photon fractional light-cone momentum
$x_{\gamma1}$ (right panel) at different energies. In the left
panel, the curves are given for fixed $x_{\gamma1}=0.5$ (solid
lines) and $x_{\gamma1}=0.9$ (dashed lines). In the right panel, the
curves are given for fixed $M^2=50\,\GeV^2$ (solid lines) and
$M^2=500\,\GeV^2$ for $\sqrt{s}=14$ TeV and 500 GeV, and
$M^2=200\,\GeV^2$ at $\sqrt{s}=40$ GeV (dashed lines). The pairs of
solid/dashed curves in the both panels correspond to
$\sqrt{s}=40\,\GeV$, $500$ GeV and $14$ TeV from top to bottom,
respectively. Here we used the KST parameterization for the
dipole-target scattering amplitude \cite{kpss,KST-GBW-eqs,KST-par}
and $F_2$ parameterization by Cudell and Soyez \cite{Cudell-F2} are
used here and below unless otherwise is specified. Unitarity
corrections are included by default into the employed
phenomenological partial elastic dipole amplitude fitted to data
(see above). In this calculation we consider the unpolarized case
summing up the contributions of longitudinal and transverse parts
both in the diffractive and inclusive cross sections.

As seen from Fig.~\ref{fig:DDYvsDY}, the DDY-to-DY cross section
ratio is falling with energy. However, naively one could expect
basing on QCD factorisation, that the DDY cross section, which is
proportional to the dipole cross section squared, should rise with
energy steeper than the total inclusive cross section. At the same
time, the ratio rises with the hard scale of the process, $M^2$.
This also looks counterintuitive, since diffraction is usually
associated with soft interactions \cite{deriv1}. These effects are
different from ones emerging in Regge factorisation-based
calculations, where we observe a slow rise of the DDY-to-DY cross
section ratio with c.m.s. energy and its practical independence on
the hard scale of the process $M^2$ \cite{Antoni11}.

In order to understand such an interesting shape of the energy and
hard scale dependence of the DDY-to-DY cross section ratio obtained
in the color dipole approach, let us look at the amplitude of the
DDY process, which is proportional to the difference between the
dipole cross sections of the Fock states with and without the hard
photon emission \cite{KPST06}, i.e.
\begin{eqnarray}
M_{DDY}\sim\sigma(\vec{R})-\sigma(\vec{R}-\alpha\vec{r})=
\frac{2\alpha\sigma_0}{R_0^2(x)}e^{R^2/R_0^2(x)}\,
\left(\vec{r}\cdot\vec{R}\right)+h.o.
\end{eqnarray}
assuming the simplest GBW slope for the dipole cross section, and
the hardness of the emitted photon implies $r\sim 1/M\ll R_0(x)$. We
see now that the diffractive DY amplitude is linear in $r$, so the
diffractive cross section turns out to be a quadratic function of $r$, which
is different from e.g. the diffractive DIS process where the cross
section is proportional to $r^4$ and is dominated by soft
fluctuations (see e.g. Refs.~\cite{KPSdiff,BP97}). Since the
diffractive DY cross section is proportional to $r^2$, then soft and
hard interactions contribute on the same footing \cite{KPST06},
which is one of the basic sources of the QCD factorisation breaking
in diffractive DY process.

As was demonstrated in Ref.~\cite{KPST06}, all the energy and scale
dependence of the DDY-to-DY cross section ratio comes via the
$x$-dependent factor,
\begin{eqnarray}
\frac{\sigma_{DDY}}{\sigma_{DY}}\propto
\frac{1}{R_0^2(x)}e^{-2R^2/R_0^2(x)}\label{ratio}
\end{eqnarray}
where $x=M^2/x_{\gamma1}s$. In our case, $R_0^2(x)<2R^2$, so the
factor in Eq.~(\ref{ratio}) rises with $R_0(x)$, i.e. with $x$. This
is the main reason why the ratio shown in Fig.~\ref{fig:DDYvsDY}
decreases with energy, but increases with the hard scale $M^2$.
Also, the falling energy behavior is partly due to the rise with
energy of the absorptive corrections \cite{KPST06}.
\begin{figure}[!h]
\begin{minipage}{0.49\textwidth}
 \centerline{\includegraphics[width=1.0\textwidth]{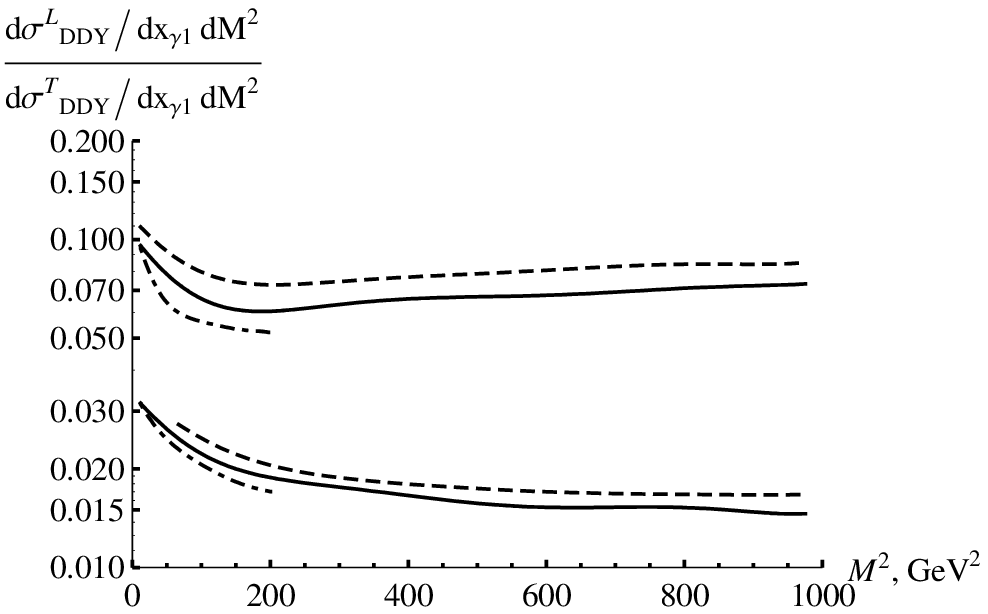}}
\end{minipage}
\hspace{0.5cm}
\begin{minipage}{0.46\textwidth}
 \centerline{\includegraphics[width=1.0\textwidth]{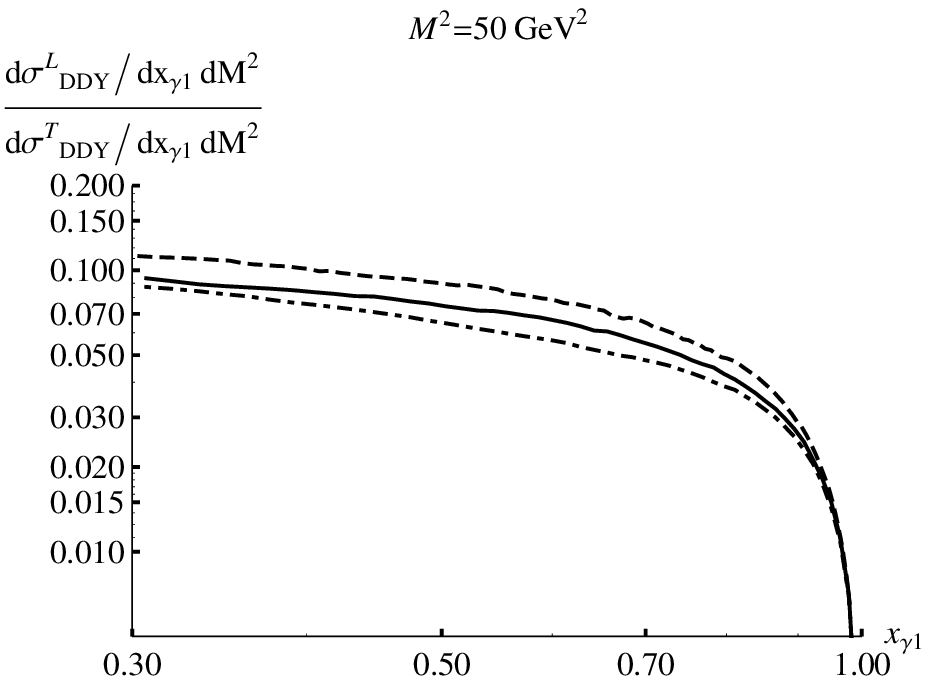}}
\end{minipage}
   \caption{
\small The ratio of the longitudinal (L) to transverse (T) photon
polarization contributions to the diffractive Drell-Yan cross
section as function of the lepton-pair invariant mass squared
$M^2$ (left panel) and photon fraction $x_{\gamma1}$ (right panel) at
different energies: $\sqrt{s}=40$ GeV (dash-dotted lines),
$\sqrt{s}=500$ GeV (solid lines) and $\sqrt{s}=14$ TeV (dashed
lines). In the left panel, the curves are given for fixed
$x_{\gamma1}=0.5$ (upper three curves) and $x_{\gamma1}=0.9$ (lower
three curves). In the right panel, the curves are given for fixed
$M^2=50\,\GeV^2$.}
 \label{fig:LvsT}
\end{figure}

In Fig.~\ref{fig:LvsT} we show the relative contribution of the
longitudinal (L) to transverse (T) photon polarization to the
diffractive Drell-Yan cross section. The ratio $\sigma_L/\sigma_T$
is presented as function of lepton-pair invariant mass squared
$M^2$ (left panel) and photon fraction $x_{\gamma1}$ (right panel) at
different energies: $\sqrt{s}=40$ GeV (dash-dotted lines),
$\sqrt{s}=500$ GeV (solid lines) and $\sqrt{s}=14$ TeV (dashed
lines). In the left panel, the curves are given for fixed
$x_{\gamma1}=0.5$ (upper three curves) and $x_{\gamma1}=0.9$ (lower
three curves). In the right panel, the curves are given for fixed
$M^2=50\,\GeV^2$. We see that the diffractive DY process is always
dominated by radiation of transversely polarized lepton pairs. The
ratio $\sigma_L/\sigma_T$ is only slightly dependent on $M^2$, and
there is no any significant energy dependence. The longitudinal
photon polarization amounts to about 10 \% at $x_{\gamma1}=0.5$ and
then steeply falls down at large $x_{\gamma1}\to1$. Such a behavior
turns out to be similar to that for inclusive DY process
\cite{KPST06}.
\begin{figure}[!h]
\begin{minipage}{0.49\textwidth}
 \centerline{\includegraphics[width=1.0\textwidth]{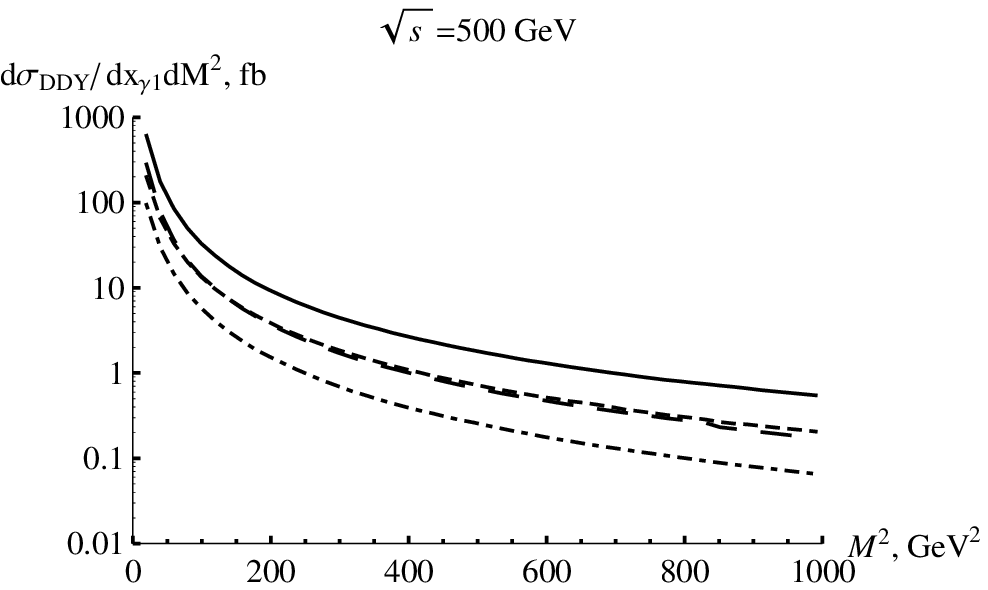}}
\end{minipage}
\hspace{0.5cm}
\begin{minipage}{0.46\textwidth}
 \centerline{\includegraphics[width=1.0\textwidth]{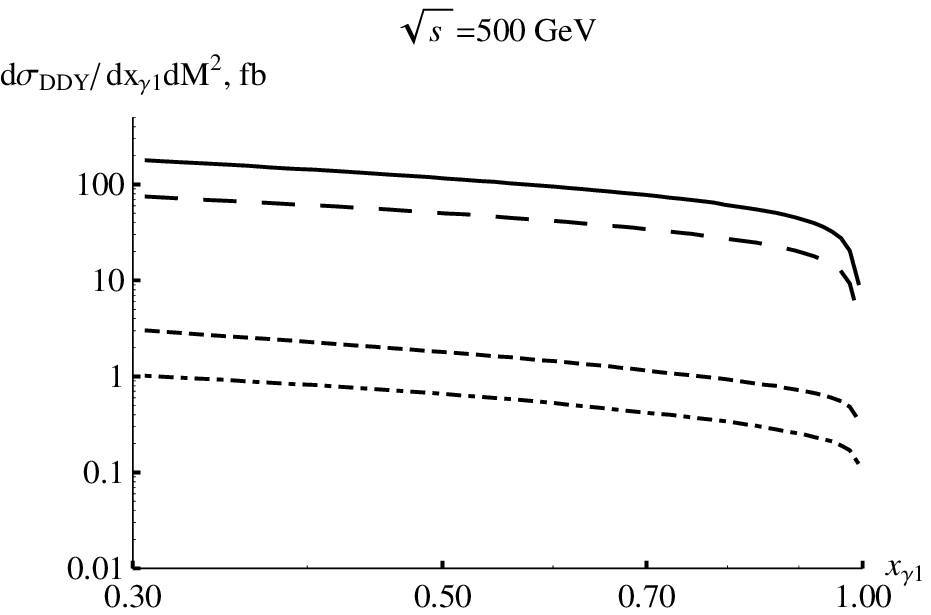}}
\end{minipage}
   \caption{
\small Diffractive Drell-Yan cross section (in fb) as function of
the lepton-pair invariant mass squared $M^2$ (left panel) and photon
fraction $x_{\gamma1}$ (right panel) at the RHIC II c.m.s. energy
$\sqrt{s}=500$ GeV. In the left panel, the curves are given for
fixed $x_{\gamma1}=0.5$ with GBW (long-dashed line) and KST (solid
line) parameterizations, and for $x_{\gamma1}=0.9$ with GBW
(dash-dotted line) and KST (dashed line) parameterizations. In the
right panel, the curves are given for fixed $M^2=50\,\GeV^2$ with
GBW (long-dashed line) and KST (solid line) parameterizations, and
for $M^2=500\,\GeV^2$ with GBW (dash-dotted line) and KST (dashed
line) parameterizations.}
 \label{fig:GBWvsKST}
\end{figure}

We also compare predictions for the diffractive DY cross section for
different parameterizations for elastic dipole-target scattering
amplitude corresponding to scattering of small (GBW given by
Eq.~(\ref{GBW})) and large (KST given by Eq.~(\ref{KST})) dipoles.
As an example, in Fig.~\ref{fig:GBWvsKST} we present the diffractive
Drell-Yan cross section as function of the lepton-pair invariant
mass squared $M^2$ (left panel) and photon fraction $x_{\gamma1}$
(right panel) at the RHIC II c.m.s. energy $\sqrt{s}=500$ GeV. We
notice that the GBW parameterization leads to roughly a factor of
two smaller cross section than the one obtained with the KST
parameterization, however, both of them exhibit basically the same
$x_{\gamma1}$ and $M^2$ shapes. It means that the evolution of the
dipole size can only affect the overall normalization of the DDY
cross section. Since arguments in the elastic amplitude $f_{el}$ in
Eq.~(\ref{amp-tot}), the impact distance between the target and the
projectile $b$ and the transverse distance between projectile quarks
$r_p\sim|\vec{r}_i-\vec{r}_j|,\,i\not=j$, are of the same order and
given at the soft hadronic scale, then the use of KST
parameterization fitted to the soft hadron scattering data data is
justified in the case of diffractive DY.
\begin{figure}[!h]
\begin{minipage}{0.49\textwidth}
 \centerline{\includegraphics[width=1.0\textwidth]{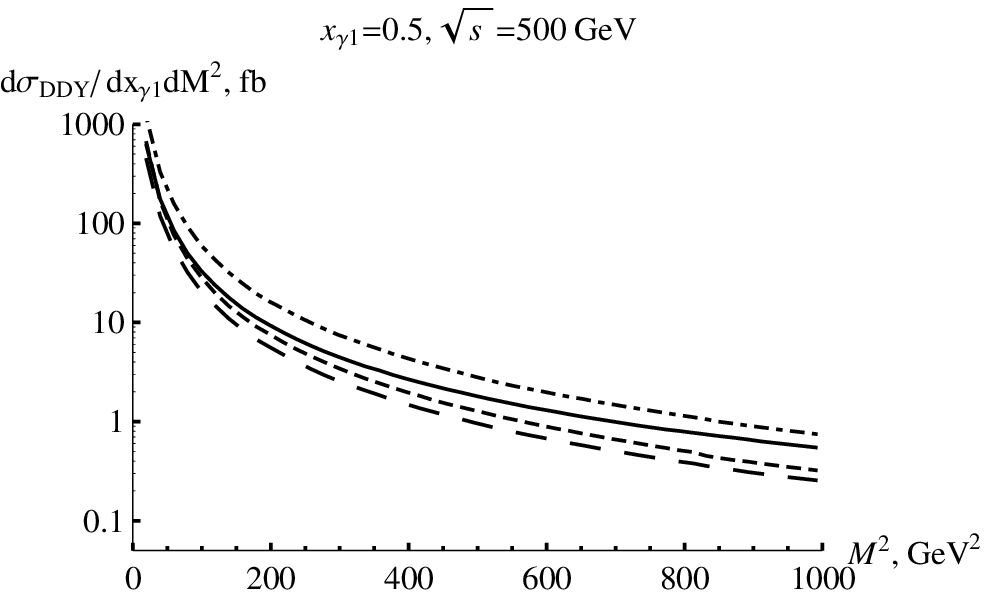}}
\end{minipage}
\hspace{0.5cm}
\begin{minipage}{0.46\textwidth}
 \centerline{\includegraphics[width=1.0\textwidth]{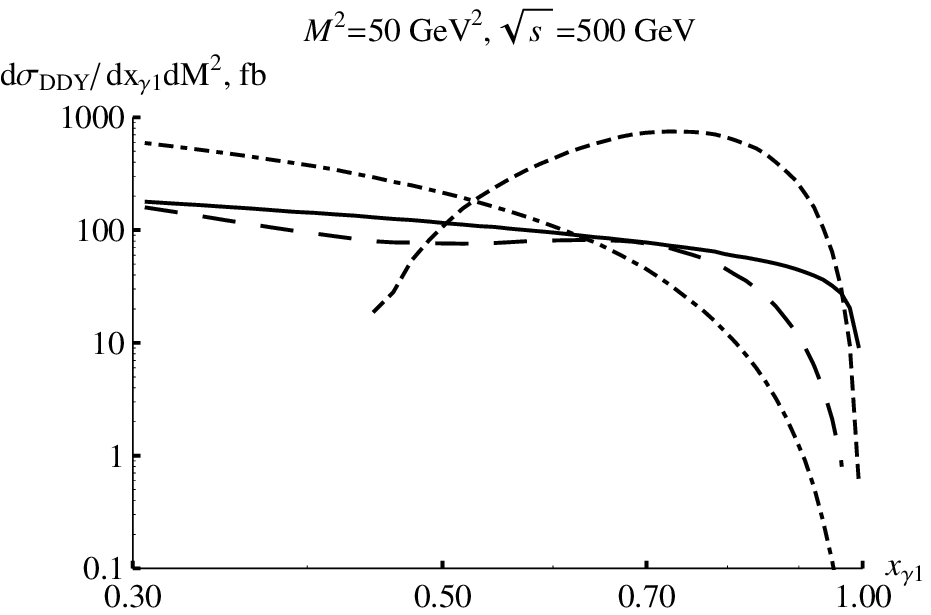}}
\end{minipage}
   \caption{
\small Diffractive Drell-Yan cross section (in fb) as function of
the lepton-pair invariant mass squared $M^2$ (left panel) with fixed
$x_{\gamma1}=0.5$ and photon fraction $x_{\gamma1}$ with fixed
$M^2=50\,\GeV^2$ (right panel) for different parameterizations of
the proton structure function $F_2$: Regge parameterization by
Cudell et al \cite{Cudell-F2} (solid line), SMC parameterization
\cite{SMC} with infrared freezing at two different scales
$Q_0^2=0.1$ (dashed line) and 0.5 $\GeV^2$ (long-dashed line) and
GJR parameterization \cite{GJR} (dash-dotted line).}
 \label{fig:F2dep}
\end{figure}

In order to illustrate the intrinsic theoretical uncertainties in
our DDY cross section calculations, in Fig.~\ref{fig:F2dep} we show
the diffractive DY cross section as function of the lepton-pair
invariant mass squared $M^2$ (left panel) and photon fraction
$x_{\gamma1}$ (right panel) for different parameterizations of the
proton structure function $F_2$ entering the DDY cross section
through Eq.~(\ref{SF}). In this figure we represent results with
four distinct cases widely used in the literature: Regge
parameterization by Cudell and Soyez \cite{Cudell-F2} (solid line),
old SMC parameterization \cite{SMC} with infrared freezing at two
different scales $Q_0^2=0.1$ (dashed line) and 0.5 $\GeV^2$
(long-dashed line) and recent GJR parameterization \cite{GJR}
(dash-dotted line). We see that the diffractive DY cross section is
sensitive to various $F_2$ parameterizations, especially at
relatively large $x_{\gamma1}\to1$, which is reflected in quite
noticeable uncertainties in our calculation. This opens up a
promising opportunity to use the diffractive DY reaction as a direct
probe of the proton structure function at rather large
$x=x_{\gamma1}/\alpha$.

The Regge parameterization \cite{Cudell-F2} is presumably
constructed in the soft region where the most of the contribution to
the DDY process comes from, and it leads to a rather regular and
stable behavior of the cross section at $x_{\gamma1}\to1$. Old SMC
$F_2$ parameterization is obviously inapplicable to the DDY calculations,
because it leads to significant uncertainties with respect to the lower
freezing scale $Q_0$ variations indicating at a strong sensitivity to
the non-perturbative low-$Q$ region where the proton structure
function $F_2$ is unknown to a large extent.

\section{Conclusion and outlook}

In this work, we have investigated in detail the QCD factorisation
breaking effects in the diffractive Drell-Yan process within the
framework of the color dipole approach. Such effects lead to quite
different properties of the corresponding observables with respect
to QCD factorisation-based calculations.

A quark cannot diffractively radiate a photon in the forward
direction, whereas a hadron can due to the presence of transverse
motion of spectator quarks in the projectile hadron. For this
reason, the diffractive DY cross section depends on the hadronic
size breaking  the QCD
factorisation.

This leads to the physical picture where hard and soft interactions
are equally important for DY diffraction, and their relative contributions are
independent of the hard scale, like in the inclusive DY process. This is a result of
the specific property of DY diffraction: its cross section is a
linear, rather than quadratic function of the dipole cross section.
On the contrary, diffractive DIS is predominantly a soft process,
because its cross section is proportional to the dipole cross
section squared.

Contrary to what follows from the calculations based on QCD
factorisation, the ratio of the diffractive to inclusive cross
sections falls with energy, but rises with the di-lepton effective
mass $M$. This happens due to the saturated behavior of the dipole
cross section which levels off at large separations. All these
properties are different from those in the diffractive DIS process,
where QCD factorisation is exact.

In addition, we made predictions for the differential (in photon
fractional momentum $x_{\gamma1}$ and di-lepton invariant mass
squared $M^2$) cross sections for the diffractive DY process at the
energies of RHIC (500 GeV) and LHC (14 TeV). The transverse photon
polarisation gives the dominant contribution to the DDY cross
section. The ratio $\sigma_L/\sigma_T$ is almost independent on
c.m.s. energy $\sqrt{s}$ and only weakly depends on the hard scale
$M^2$.

Finally, we propose an alternative treatment of the absorptive
effects describing the subsequent soft interactions of projectile
quarks off the proton target by multiple elastic dipole-target
scatterings. Such an idea leads to eikonalization of the ``bare''
elastic dipole amplitude in the DDY amplitude and, ultimately, to a
description of the gap survival effects on the same footing with the
leading-order diffractive DY subprocess in the framework of the
color dipole approach without introducing the gap survival factor
$K$. These corrections are included by default into the employed
phenomenological partial elastic dipole amplitude fitted to data.

The main features of the diffractive Drell-Yan reaction described
above are valid for other diffractive Abelian processes, like
production of direct photons, Higgsstrahlung, radiation of $Z$ and
$W$ bosons. A detailed analysis of these processes in the framework
of the color dipole approach is planned for a forthcoming study.

{\bf Acknowledgments}

Useful discussions and helpful correspondence with Jochen Bartels,
Antoni Szczurek, Gunnar Ingelman and Mark Strikman are gratefully
acknowledged. This study was partially supported by the Carl Trygger
Foundation (Sweden), by Fondecyt (Chile) grant 1090291, and by
Conicyt-DFG grant No. 084-2009. Authors are also indebted to the
Galileo Galilei Institute of Theoretical Physics (Florence, Italy)
and to the INFN for partial support and warm hospitality during
completion of this work.


\end{document}